\documentclass[preprint,preprintnumbers,amsmath,amssymb,superscriptaddress]{revtex4}
\usepackage{txfonts}
\usepackage{graphicx}
\usepackage{dcolumn}
\usepackage{bm}
\usepackage{array}
\usepackage[final]{changes}
\begin{document}

\title{Structural phase transition, precursory electronic anomaly and strong-coupling superconductivity in quasi-skutterudite (Sr$_{1-x}$Ca$_{x}$)$_{3}$Ir$_{4}$Sn$_{13}$ and Ca$_{3}$Rh$_{4}$Sn$_{13}$\thanks{Project supported by the National Natural Science Foundation of China (No. 11674377, No. 11634015), the National Key R$\&$D Program of China (No. 2017YFA0302904 and No. 2016YFA0300502), and the Strategic Priority Research Program(B) of the Chinese Academy of Sciences (No. XDB07020200).}}


\author{Jun Luo$^{1,2}$, \ Jie Yang$^{1}$, \ S. Maeda$^{3}$, \ Zheng Li$^{1,2}$, \ and \ Guo-Qing Zheng$^{1,2,3}$\\
\small $^{1}${Institute of Physics, Chinese Academy of Sciences, and Beijing National Laboratory for Condensed Matter Physics,Beijing 100190, China}\\  
\small $^{2}${School of Physical Sciences, University of Chinese Academy of Sciences, Beijing 100049, China}\\
\small $^{3}${Department of Physics, Okayama University, Okayama 700-8530, Japan}}
\date{\today}

\begin{abstract}
The interplay between superconductivity and structural phase transition has attracted enormous interests in recent years. For example, in Fe-pnictide high temperature superconductors, quantum fluctuations in association with structural phase transition have been proposed to lead to many novel physical properties and even the superconductivity itself. Here we report a finding that the quasi-skutterudite superconductors  (Sr$_{1-x}$Ca$_{x}$)$_{3}$Ir$_{4}$Sn$_{13}$ ($x$ = 0, 0.5, 1) and Ca$_{3}$Rh$_{4}$Sn$_{13}$ show some unusual properties similar to the Fe-pnictides, through $^{119}$Sn nuclear magnetic resonance (NMR) measurements. In (Sr$_{1-x}$Ca$_{x}$)$_{3}$Ir$_{4}$Sn$_{13}$, the NMR linewidth increases below a temperature $T^*$ that is higher than the structural phase transition temperature $T_{\rm s}$. The spin-lattice relaxation rate ($1/T_1$) divided by temperature ($T$), $1/T_1T$, and the Knight shift $K$ increase with decreasing $T$ down to $T^*$, 
but start to decrease below $T^*$ and followed by  more distinct changes at $T_{\rm s}$.
In contrast, none of the anomalies was observed in Ca$_{3}$Rh$_{4}$Sn$_{13}$ that does not undergo a structural phase transition. The precursory phenomenon above structural phase transition resembles that occurs in Fe-pnictides.
In the superconducting state of Ca$_{3}$Ir$_{4}$Sn$_{13}$, $1/T_{1}$ decays as ${\rm exp}(-\Delta/k_{\rm B}T)$ with a large gap $\Delta = 2.21 k_{\rm B}T_{\rm c}$, yet without a Hebel-Slichter coherence peak, which indicate strong-coupling superconductivity.
Our results provide new insight into the relationship between superconductivity and the electronic-structure change associated with structural phase transition.
\end{abstract}


\maketitle

\section{Introduction}
Transition-metal compounds show diverse properties such as magnetism, superconductivity and charge density wave, and often accompany a structural transition.\cite{TaSe2,NaCoO,FeAs} In these materials, the interplay between superconductivity  and other orders is of great interests. 
For example, in the copper oxides,\cite{cuprates} heavy fermions,\cite{Lonzarich} and iron-based
superconductors\cite{Oka} that contain transition metal elements, superconductivity is found in the vicinity of a quantum critical point (QCP) at which other orders are completely suppressed  at absolute zero temperature. In particular, in iron-based superconductors, not only a magnetic (spin density wave) QCP, but also another QCP associated with the structural phase transition  exists.\cite{Zhou} In this case, quantum fluctuations of electronic nematic order associated with the structural phase transition may lead to many novel physical properties such as $T$-linear electrical resistivity.\cite{Zhou,Yang,Fisher,Kivelson}

Materials with the general stoichiometry R$_{3}$M$_{4}$X$_{13}$ are a large family usually adopting a common
quasi-skutterudite structure, where R is alkaline-earth or rare-earth element, M is a transition metal and X is a group-IV element.\cite{R3M4X13,akimitsu} Superconductivity with a fairly high transition temperature $T_{\rm c}$ $\sim$ 7 K has been found in R$_{3}$M$_{4}$Sn$_{13}$ more than 30 years ago,\cite{crystalgrowth,crystal} but the physical properties are poorly known.
Recently, this class of materials received new attention because of a possible interplay between the
superconductivity and the structure instability.

The electrical resistivity, susceptibility, Hall coefficient and heat transport measurements on
Ca$_{3}$Ir$_{4}$Sn$_{13}$ found that an anomaly occurs at a   temperature of 35 K, above the superconducting
transition temperature $T_{\rm c}$ = 7 K.\cite{yangjh,Lisy,Hall} The anomaly was ascribed to ferromagnetic spin fluctuation in early works.\cite{yangjh} Resistivity and susceptibility measurements on Sr$_{3}$Ir$_{4}$Sn$_{13}$ also show anomaly at 147 K. Subsequent X-ray diffraction and pressure effect measurements on Sr$_{3}$Ir$_{4}$Sn$_{13}$ showed that a structural phase transition from a cubic $I$ phase ($Pm\overline{3}n$) to a $I^{'}$ phase ($I\overline{4}3d$) takes place at $T_{\rm s}$, with the lattice parameter doubled in the low temperature phase. Therefore, the anomalies reported earlier in this class of materials are due to the structural phase transition. By chemical or physical pressure, $T_{\rm s}$ can be suppressed to zero, while $T_{\rm c}$ increases slowly and reaches to a maximum 8.9 K.\cite{Grosche1} Similar phase diagram has also been obtained in (Sr$_{1-x}$Ca$_{x}$)$_{3}$Rh$_{4}$Sn$_{13}$.\cite{Grosche2} Thus, a possible relation between the enhanced $T_{\rm c}$ and the structure instability  has been suggested.\cite{Grosche1,Goh}

The nature of the electronic structure change due to the structural phase transition is not well understood.
Neutron scattering and specific heat measurements reveal a second-order nature of the structural phase
transition in (Sr$_{1-x}$Ca$_{x}$)$_{3}$Ir$_{4}$Sn$_{13}$.\cite{Neutron,themodynamics} Hall coefficient changed from a negative to a positive value and the optical measurement indicated that the Drude spectral weight is transferred to the high energy region across $T_{\rm s}$ in Sr$_{3}$Ir$_{4}$Sn$_{13}$.\cite{wangnl,NMR2} Based on these results, a reconstruction of Fermi surface below $T_{\rm s}$ due to a charge density wave (CDW) formation was suggested.\cite{wangnl,NMR2}

In this work, we grow single crystals of  (Sr$_{1-x}$Ca$_{x}$)$_{3}$Ir$_{4}$Sn$_{13}$ ($x$ = 0, 0.5, 1) and
Ca$_{3}$Rh$_{4}$Sn$_{13}$, and perform electrical resistivity and $^{119}$Sn NMR measurements to elucidate the
electronic properties change associated with the structural phase transition.
(Sr$_{1-x}$Ca$_{x}$)$_{3}$Ir$_{4}$Sn$_{13}$ ($x$ = 0, 0.5, 1) undergo a structural phase transition at $T_{\rm s}$ = 147 K, 85 K and 35 K respectively, while  Ca$_{3}$Rh$_{4}$Sn$_{13}$ does not. By NMR measurements, we find that an anomaly occurs already above $T_{\rm s}$  in (Sr$_{1-x}$Ca$_{x}$)$_{3}$Ir$_{4}$Sn$_{13}$ ($x$ = 0, 0.5, 1). 
Such electronic anomaly prior to structural transition resembles an actively-investigated phenomenon in some of the Fe-based superconductors where  the physical properties show an in-plane anisotropy (nematicity) above $T_{\rm s}$ below which the C$_4$ symmetry is lowered to C$_2$ symmetry. 
However, in Ca$_{3}$Rh$_{4}$Sn$_{13}$ that does not undergo a structural transition, the Korringa relation is
satisfied down to $T \sim$ 20 K. The electronic state properties below $T^*$ are discussed by analyzing the change in the Korringa ratio.
We also measured the superconducting state property of Ca$_{3}$Ir$_{4}$Sn$_{13}$, and found that it is  a
strong-coupling s-wave superconductor. We will discuss the relationship between superconductivity,  the electronic state change associated with the structural transition and  electron correlations.
\section{Experiment}
\noindent Single crystals of (Sr$_{1-x}$Ca$_{x}$)$_{3}$Ir$_{4}$Sn$_{13}$, Ca$_{3}$Rh$_{4}$Sn$_{13}$ were grown by
self-flux method, as previously reported in Ref.~\cite{crystalgrowth}. The composition shown in this paper is the
nominal one. Excessive Sn flux was removed in concentrated HCl acid. Crystals with proper size were picked up and
polished, then the temperature dependence of resistivity were measured by standard four-probe method using physical
properties measurement system (PPMS, Quantum Design). The $T_{\rm c}$ was determined by both DC susceptibility using
Magnetic properties measurement system(MPMS, Quantum Design) with an applied magnetic field of 10 Oe, and AC
susceptibility using an $in-situ$ NMR coil. For $^{119}$Sn NMR measurements, since the sample shows a  good
electrical conductivity so that the skin depth is short, we crushed the single crystals into fine powders to gain
the surface area. 
The $^{119}$Sn nucleus has a nuclear spin $I$ = 1/2 and gyromagnetic ratio $\gamma_{\rm n}$/$2\pi$ = 15.867 MHz/T.
The $^{119}$Sn NMR spectra were obtained by scanning rf frequency and integrating spin echo intensity at a fixed
magnetic field $H_{0}$. The spin-lattice relaxation time $T_{1}$ was measured by using the saturation-recovery
method, and obtained by a good fitting of the nuclear magnetization $M(t)$  to $1-M(t)/M_{0} = \exp(-t/T_{1})$,
where $M(t)$ is the nuclear magnetization at time $t$ after the single saturation pulse and $M_0$ is the nuclear
magnetization at thermal equilibrium.

\section{Results}

\subsection{Sample characterization}
\begin{figure}[h]
\includegraphics[width= 7cm]{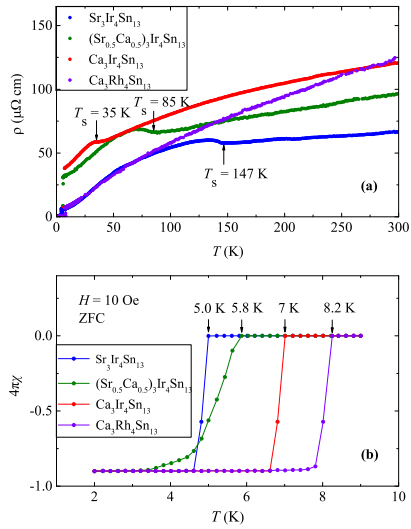}
\centering
\caption{(Color online) 
(a) The temperature dependence of the electrical resistivity for (Sr$_{1-x}$Ca$_{x}$)$_{3}$Ir$_{4}$Sn$_{13}$ ($x$ = 0, 0.5, 1) and Ca$_{3}$Rh$_{4}$Sn$_{13}$. The arrows indicate the structural transition temperature $T_{\rm s}$. (b) The temperature dependence of the magnetic susceptibility for (Sr$_{1-x}$Ca$_{x}$)$_{3}$Ir$_{4}$Sn$_{13}$ ($x$ = 0, 0.5, 1) and Ca$_{3}$Rh$_{4}$Sn$_{13}$ below \emph{T} = 9 K. The arrows indicate the critical temperature $T_{\rm c}$.
\label{RTsusceptibility}}
\end{figure}
Figure~\ref{RTsusceptibility}.(a) presents the temperature dependence of the electrical resistivity for
(Sr$_{1-x}$Ca$_{x}$)$_{3}$Ir$_{4}$Sn$_{13}$ ($x$ = 0, 0.5, 1) and Ca$_{3}$Rh$_{4}$Sn$_{13}$ single crystals. For (Sr$_{1-x}$Ca$_{x}$)$_{3}$Ir$_{4}$Sn$_{13}$ ($x$ = 0, 0.5, 1) the resistivity shows a distinct hump, which has been ascribed to structure transition.\cite{Grosche1} Figure~\ref{RTsusceptibility}.(b) shows the temperature dependence of DC magnetic susceptibility below 9 K. A sharp superconducting transition with a narrow transition width of about 0.2 K for Sr$_{3}$Ir$_{4}$Sn$_{13}$, Ca$_{3}$Ir$_{4}$Sn$_{13}$ and Ca$_{3}$Rh$_{4}$Sn$_{13}$, which indicates a good crystal quality. 
The $T_{\rm c}$ is determined by the point at which the magnetic susceptibility begins to decrease.  The Meissner shielding fraction 
is estimated to be over 90\%, which proves a bulk nature of superconductivity.
\subsection{Precursory electronic anomaly above  $T_{\rm s}$}
\begin{figure}[h]
\includegraphics[width= 7cm]{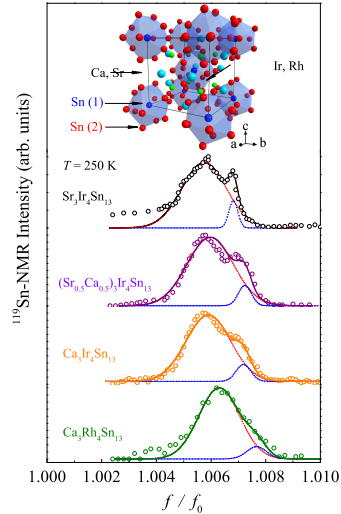}
\centering
\caption{(Color online) 
The crystal structure of (Sr$_{1-x}$Ca$_{x}$)$_{3}$Ir$_{4}$Sn$_{13}$  and Ca$_{3}$Rh$_{4}$Sn$_{13}$ that contain  two different Sn sites, and the $^{119}$Sn-NMR spectra for (Sr$_{1-x}$Ca$_{x}$)$_{3}$Ir$_{4}$Sn$_{13}$ ($x$ = 0, 0.5, 1) and Ca$_{3}$Rh$_{4}$Sn$_{13}$ at 250 K.
The horizontal coordinate-axis  is the reduced frequency, $f/f_0$, with $f_0 = \gamma H_0/2\pi$.
The blue and red dotted lines are Gaussian fittings to the obtained spectra with area ratio of 1 : 12. The solid lines are the sum of the two Gaussian functions.
\label{spec250K}}
\end{figure}

Since $^{119}$Sn ($I$ = 1/2) has no quadrupole moment, the nuclear spin Hamiltonian is simply given
by the Zeeman interaction, $\mathcal{H}= -\gamma_{\rm n}\hbar H_{0}(1+K)\hat{I}$, where $K$ is the Knight shift and $\gamma_{\rm n}$ is the nuclear gyromagnetic ratio. As expected  in a material with a cubic crystal structure, a single $^{119}$Sn NMR transition line (m = -1/2 $\leftrightarrow$ m = 1/2 transition) was observed. Considering that the Sn atoms form an icosahedral cage in the crystal structure and there are two different crystallographic sites (See Fig.~\ref{spec250K}), namely one Sn(1) in the center and twelve Sn(2) on the vertices of icosahedron, the spectrum should have two peaks. Figure~\ref{spec250K} shows the frequency-swept $^{119}$Sn-NMR spectra measured at $T$ = 250 K under a fixed magnetic field for the four samples. It can be seen that the spectra indeed show two peaks.
All the spectra can be fitted by two Gaussian functions with an area ratio of 12 : 1, so the low frequency peak and high frequency peak respectively correspond to Sn(2) and Sn(1), which have an occupancy ratio of 12 : 1. The typical fitting curves for 250 K are shown in Fig.~\ref{spec250K}.

\begin{figure}[h]
\includegraphics[width= 15cm]{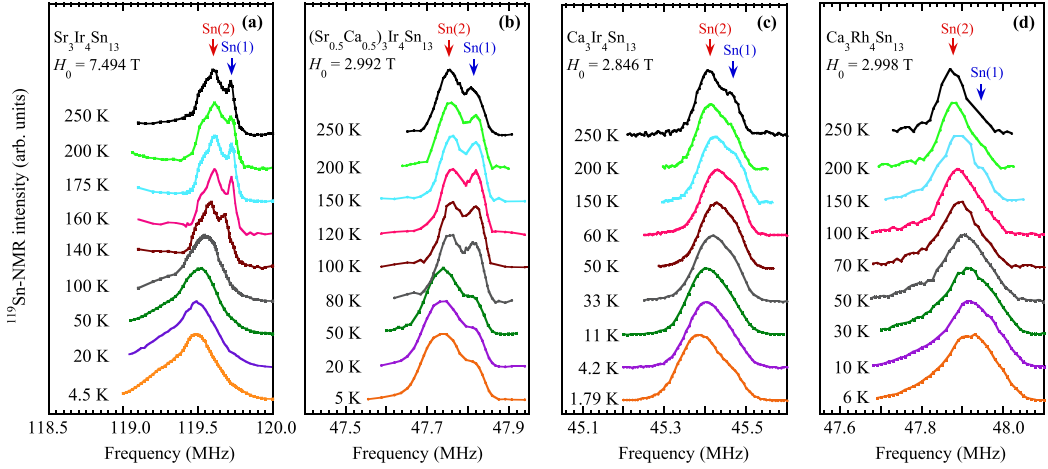}
\centering
\caption{(Color online) 
The temperature dependence of the $^{119}$Sn-NMR spectra for (Sr$_{1-x}$Ca$_{x}$)$_{3}$Ir$_{4}$Sn$_{13}$ ($x$ = 0, 0.5, 1) and Ca$_{3}$Rh$_{4}$Sn$_{13}$.
The two peaks marked by blue arrow and red arrow correspond to Sn(1) site and Sn(2) site, respectively.
\label{specfoursamples2}}
\end{figure}

\begin{figure}[h]
\includegraphics[width= 15cm]{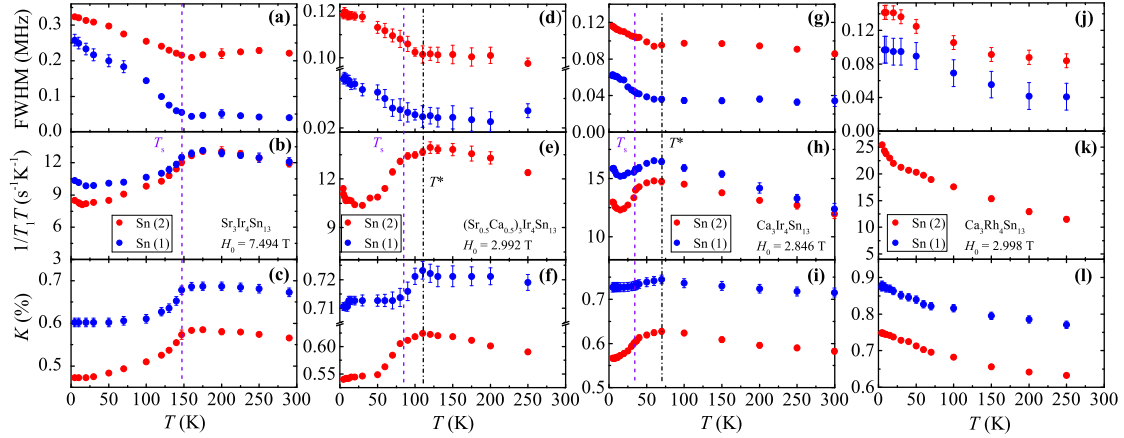}
\centering
\caption{(Color online) 
The temperature dependence of FWHM of NMR spectra, $1/T_{1}T$ and the Knight shift $K$  of (Sr$_{1-x}$Ca$_{x}$)$_{3}$Ir$_{4}$Sn$_{13}$ ($x$ = 0, 0.5, 1) and Ca$_{3}$Rh$_{4}$Sn$_{13}$. The blue and red filled circle correspond to Sn(1) and Sn(2) site, respectively. The purple dash lines indicate structural transition temperature $T_{\rm s}$. The black dash dot lines indicate the temperature $T^*$ below which  anomalies begin.
\label{T1TKFWHM}}
\end{figure}

Below we discuss the normal-state properties inferred from the NMR measurements.
Figure~\ref{specfoursamples2} shows the temperature dependence of spectra for the four samples, from which the full width at half maximum (FWHM) of spectra is obtained as shown in Fig.~\ref{T1TKFWHM}.
Usually, one expects that the   FWHM of the spectra increases  below  a structural phase transition temperature, below which  four types of Sn(2) sites are formed\cite{Grosche1} which may have slightly different $K$ and result in broadening of spectra.
However, as can be seen in the Fig.~\ref{T1TKFWHM}, the FWHM starts to increase at a temperature $T^*$ that is above $T_{\rm s}$.

Figure~\ref{T1TKFWHM} \added{also} shows the temperature dependence of $1/T_{1}T$ and the Knight shift $K$ for the four samples. The $K$ was obtained from the Gaussian fitting of the spectra. $T_{1}$ was measured at the position of the respective two peak. It can be seen that  $1/T_{1}T$ and $K$ also decrease below $T^*$, and followed by a more distinct change at $T_{\rm s}$. \added{For (Sr$_{0.5}$Ca$_{0.5}$)$_{3}$Ir$_{4}$Sn$_{13}$ and
Ca$_{3}$Ir$_{4}$Sn$_{13}$, the anomaly is pronounced, although it is less clear for Sr$_{3}$Ir$_{4}$Sn$_{13}$.} A previous report in Ca$_{3}$Ir$_{4}$Sn$_{13}$ that $1/T_{1}T$ decreases below around 75 K is consistent with our results.\cite{ChenB}

The total Knight shift consists of three parts,
$K = K_{\rm dia} + K_{\rm orb} + K_{\rm s}$,
where $K_{\rm dia}$ arises from the diamagnetic susceptibility $\chi_{\rm dia}$, $K_{\rm orb}$ from the orbital
(Van-Vleck) susceptibility $\chi_{\rm orb}$,  and $K_{\rm s}$ from the spin susceptibility $\chi_{\rm s}$,
respectively.
The $\chi_{\rm s}$ is estimated to be 8.1$\times$10$^{-4}$emu/mol according to $\chi_{\rm s}$ = 2$\mu_{\rm
B}^2$$N(0)$, where $N(0)$ = 12.5 eV$^{-1}$ per formula unit is the density of states at Fermi
level.\cite{electronicstrcutrue}
The closed shells or fully occupied electronic bands contribute to the diamagnetism, which is proportional to the atomic number and the atomic radius. Since iridium has a large atomic number, the reported diamagnetic
susceptibility of Sr$_{3}$Ir$_{4}$Sn$_{13}$, which is -1.2$\times$10$^{-4}$emu/mol above $T_{\rm s}$,\cite{Grosche1,wangnl} is mainly due to iridium, which is -9.8$\times$10$^{-4}$emu/mol.\cite{yangjh} The
$\chi_{\rm orb}$ is temperature-independent, and is usually much smaller than $\chi_{\rm s}$. After considering
different contributions, the total susceptibility can be diamagnetic as reported.\cite{Grosche1,wangnl}
However, the Ir diamagnetism has no hyperfine coupling to Sn nuclear spins. Thus the Sn Knight shift is mainly due to $\chi_{\rm s}$, and is positive, as found in our measurement.

In genaral, $1/T_{1}T$ probes the transverse imaginary part of the dynamic susceptibility ($\chi''_\perp(q,\omega)$)
and can be written as
\begin{equation}
\frac{1}{T_1T} = \frac{2{\gamma_{\rm n}}^2k_ {\rm B}}{(\gamma_{\rm e}\hbar)^2}\sum_q |A(q)|^2
\frac{\chi''_\perp(q,\omega)}{\omega},
\end{equation}
where $A(q)$ is the hyperfine coupling constant, $\omega$ is the NMR frequency, and $\gamma_{\rm e}$ is the
electronic gyromagnetic ratio. In a simple metal with no electron correlation, $1/T_{1}T$ is reduced to a constant proportional to $N(0)^2$. 
On the other hand, $K_{\rm s}$ due to spin susceptibility is proportional to $N(0)$. As a result, a relation between the two quantities (Korringa relation), $T_{1}TK_{\rm s}^2$ = $\frac{\hbar}{4\pi k_ {\rm B}}$$(\frac{\gamma_{\rm e}}{\gamma_{\rm n}})^2$, is obtained. Therefore, a decrease of $1/T_{1}T$ and $K$ below $T_{\rm s}$ is usually encountered, as  a reduction  of the density of states (DOS) can be expected. 
A closer look into the data finds that 
$1/T_{1}T$ and $K$ decrease more rapidly at Sn(2) site than Sn(1) site. This indicates that  the Sn(2) site has a more close relationship with the transition. This is consistent with the Neutron and X-ray scttering experiments that have found a breathing mode of phonon due to Sn(2) atoms which is softened at  $T_{\rm s}$.\cite{Neutron} In contrast to (Sr$_{1-x}$Ca$_{x}$)$_{3}$Ir$_{4}$Sn$_{13}$ ($x$ = 0, 0.5, 1), we didn't detect any anomaly in temperature dependance of $1/T_{1}T$, $K$ and FWHM of Ca$_{3}$Rh$_{4}$Sn$_{13}$ above 20 K , as shown in Fig.~\ref{T1TKFWHM} (j) (k) (l). This is consistent with the fact that Ca$_{3}$Rh$_{4}$Sn$_{13}$ does not undergo structural phase transition.

The anomaly seen in $1/T_{1}T$ and  $K$ at $T^*$ resembles the puzzling phenomenon  of the Fe-based superconductors such as BaFe$_{2-x}$Co$_x$As$_2$, BaFe$_2$(As$_{1-x}$P$_x$)$_2$ or NaFe$_{1-x}$Co$_x$As,\cite{BaFeAsP,BaFeCoAs,Zhou2} where nematic properties already appear at a temperature  far above $T_{\rm s}$. In this class of Fe-based materials, a $\rm C_{4}$ to $\rm C_{2}$ structural phase transition \replaced{takes}{take} place at $T_{\rm s}$.
It also shares some similarities with the pseudogap behaviors in underdoped copper oxide superconductors, where the DOS starts to decrease before superconducting phase transition.\cite{pseudogap}
In any event, the precursory  electronic anomaly above the $T_{\rm s}$ suggests that the structural phase transition is electronically-driven, rather than lattice-driven.
\begin{figure}[h]
\includegraphics[width= 14cm]{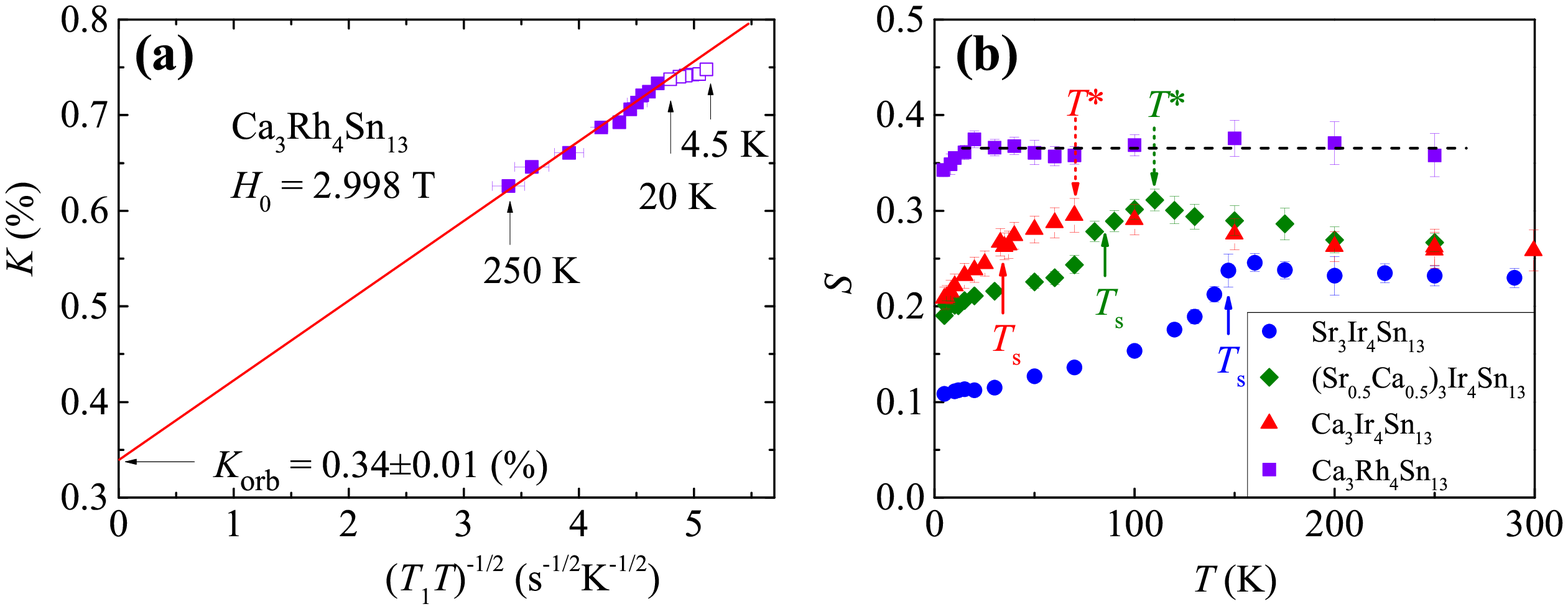}
\centering
\caption{(Color online) 
(a) The relationship between $(T_{1}T)^{-1/2}$ and $K$ for Ca$_{3}$Rh$_{4}$Sn$_{13}$, with temperature as an implicit parameter. The red solid line is a linear fitting to the data above $T$ = 20 K. The solid violet squares and open violet squares represent points above 20 K and below 20K, respectively. (b) The Korringa ratio as function of temperature  for (Sr$_{1-x}$Ca$_{x}$)$_{3}$Ir$_{4}$Sn$_{13}$ ($x$ = 0, 0.5, 1) and Ca$_{3}$Rh$_{4}$Sn$_{13}$.  The dash arrows indicate $T^*$. The solid arrows indicate $T_{\rm s}$. The  dashed straight line is the guide to the eyes.
\label{CaRhSnKorbKorringa}}
\end{figure}
\subsection{Electronic-state change below \replaced{$T^*$}{$T_{\rm s}$}}
Next, we examine if the electronic-state change below \replaced{$T^*$}{$T_{\rm s}$} can  be totally ascribed to a loss of the DOS.
In Fig.~\ref{CaRhSnKorbKorringa} (a), we plotted  $K$ against $(T_{1}T)^{-1/2}$ for Ca$_{3}$Rh$_{4}$Sn$_{13}$. In the temperature range 20 K $<$ \emph{T} $<$ 250 K, a good linear relation was found between   $K$ and
$(T_{1}T)^{-1/2}$, as expected for a conventional metal where both $(T_{1}T)^{-1/2}$ and $K_{\rm s}$ are
proportional to the DOS. Deviation from the linear relation below $T$ = 20 K will be discussed later. From the intercept of the relation, we obtain $K_{\rm orb}$ = 0.34$ \pm $0.01$\%$.

For (Sr$_{1-x}$Ca$_{x}$)$_{3}$Ir$_{4}$Sn$_{13}$ ($x$ = 0, 0.5, 1), however, the temperature dependence of
$(T_{1}T)^{-1/2}$ and $K$ is weak at  high temperatures so that a similar plot does not yield  meaningful
information. Instead we plot  the so-called Korringa ratio $S$ as a function of temperature in Fig.~\ref{CaRhSnKorbKorringa} (b), where $S$ is defined as
\begin{equation}
S = \frac{4\pi k_ {\rm B} T_1 T K_{\rm s}^2}{\hbar} (\frac{\gamma_{\rm n}}{\gamma_{\rm e}})^2
\end{equation}
For a conventional metal, $S$ = 1.
As one can see, $S$ is constant for Ca$_{3}$Rh$_{4}$Sn$_{13}$ above $T$ = 20 K within the experimental error, which is also true for \deleted{Sr$_{3}$Ir$_{4}$Sn$_{13}$} \added{(Sr$_{1-x}$Ca$_{x}$)$_{3}$Ir$_{4}$Sn$_{13}$} above $T^*$. 
However, $S$ shows a distinct decrease below $T^*$ for (Sr$_{1-x}$Ca$_{x}$)$_{3}$Ir$_{4}$Sn$_{13}$, which suggests that the reduction of \replaced{$1/T_{1}T$}{$(1/T_{1}T)$} and $K$ cannot simply be ascribed to a loss of the DOS. Below, we discuss possibilities for the decrease of $S$.

Firstly, a second-order phase transition often accompanies with the development of a short-range correlation just above the transition temperature, thus the structural instability may be responsible for the anomaly seen in our NMR data. Theoretical calculations of phonon dispersion suggest that imaginary phonon modes exist in Sr$_{3}$Ir$_{4}$Sn$_{13}$ and the lattice instabilities lie at some wave vectors.\cite{electronicstrcutrue} Neutron scattering data have reported the softening of phonon mode towards $T_{\rm s}$.\cite{Neutron} Specific heat measurements on Sr$_{3}$Ir$_{4}$Sn$_{13}$ also show that $\Delta$$C/T$ starts to increase at 160 K ($T_{\rm s}$ = 147 K) and critical fluctuation model can fit the specific heat data well, which leads to the proposal of short-range correlation above $T_{\rm s}$.\cite{themodynamics}
Therefore, the NMR quantities may also be affected by such structural short-range correlation through
magneto-elastic coupling, resulting in the deviation from the  Korringa relation below $T^*$. On the other hand, we note that, for a CDW case, the quantity $1/T_{1}T$ will increase with decreasing temperature towards the transition temperature,\cite{Kawasaki,lizheng} in contrast to a decrease of $1/T_{1}T$ observed in the present case.

\begin{figure}[h]
\includegraphics[width= 9cm]{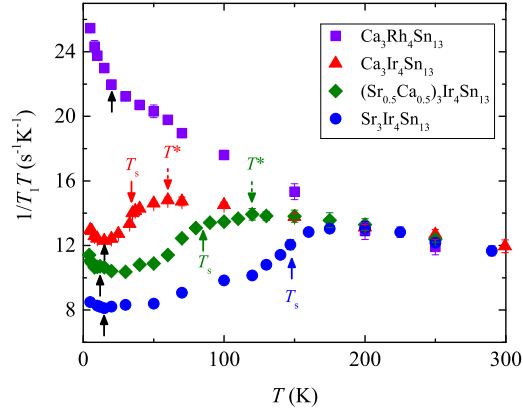}
\centering
\caption{(Color online) 
The temperature dependence of $1/T_{1}T$ for (Sr$_{1-x}$Ca$_{x}$)$_{3}$Ir$_{4}$Sn$_{13}$ ($x$ = 0, 0.5, 1) and Ca$_{3}$Rh$_{4}$Sn$_{13}$. The dash arrows indicate $T^*$. The solid arrows indicate $T_{\rm s}$. The solid black arrows mark the temperature below which $1/T_{1}T$ shows an upturn.
\label{T1Tfoursamples1}}
\end{figure}
Secondly, we discuss the possibility of magnetic correlations. To explore this issue in more detail, we turn to the data at temperatures below $T$ = 20 K. \deleted{where the Knight shift is $T$-independent for
(Sr$_{1-x}$Ca$_{x}$)$_{3}$Ir$_{4}$Sn$_{13}$. }In Fig.~\ref{T1Tfoursamples1}, we compare $1/T_{1}T$ of the four
samples. Above 160 K, $1/T_{1}T$ for all samples has  the same value and shows a similar temperature variation.
Below 160 K, $1/T_{1}T$ of Ca$_{3}$Rh$_{4}$Sn$_{13}$ increases all the way down to  4.2 K, while $1/T_{1}T$ of
(Sr$_{1-x}$Ca$_{x}$)$_{3}$Ir$_{4}$Sn$_{13}$ is reduced below \replaced {$T^*$}{$T_{\rm s}$}.
Interestingly, we note that, in all samples, there exists an upturn at a low temperature below $T$ = 20 K, as
indicated by the \added{black} arrows. Such upturn is quite pronounced particularly in Ca$_{3}$Rh$_{4}$Sn$_{13}$. \added{It can be seen from Fig. \ref{T1TKFWHM} that $1/T_{1}T$ show a clear upturn for (Sr$_{1-x}$Ca$_{x}$)$_{3}$Ir$_{4}$Sn$_{13}$, while the Knight shift is $T$-independent in such
temperature range. For  Ca$_{3}$Rh$_{4}$Sn$_{13}$, the temperature of $1/T_{1}T$ and $K$ also deviate from Korringa relation below $T$ = 20 K as due to the additional increase of $1/T_{1}T$.} Therefore, the upturn is clearly not due to a change in the DOS \added{or the structural instability, since the upturn is away from $T_{\rm s}$ and Ca$_{3}$Rh$_{4}$Sn$_{13}$ even does not undergo a structural phase transition.}
Previously, anharmonic phonons due to the rattling motion of the ions inside the cage
have been proposed to contribute an additional relaxation,\cite{Ueda}
and possibly have been seen in filled-skutterudites LaOs$_{4}$Sb$_{12}$ and
LaPt$_{4}$Ge$_{12}$.\cite{LaOsSb,LaPtGe}
In the \replaced{present}{presents} systems, one might also expect that the Sn(1) atoms rattle inside the cage and additionally contribute to $1/T_{1}T$. However our data show that Sn(1) and Sn(2) sites exhibit the same behavior, even though the Sn(2) atoms are not involved in the rattling motion.  Therefore,  the possibility of rattling as a cause for the upturn in $1/T_{1}T$ may be excluded.
Rather,  the rise of $1/T_1T$ at low temperatures  likely originates from antiferromagnetic spin fluctuations cause an increase of $\chi''_\perp(q,\omega)$ at a finite $q$ when the temperature is lowered. Therefore, coming back to Fig.~\ref{CaRhSnKorbKorringa} (b), \added{we believe that,}  the reduction of  $S$ is more likely due to antiferromagnetic spin fluctuations that develop below $T^*$ and become more evident at low temperatures.
\subsection{Phase diagram }
\begin{figure}[h]
\includegraphics[width= 7cm]{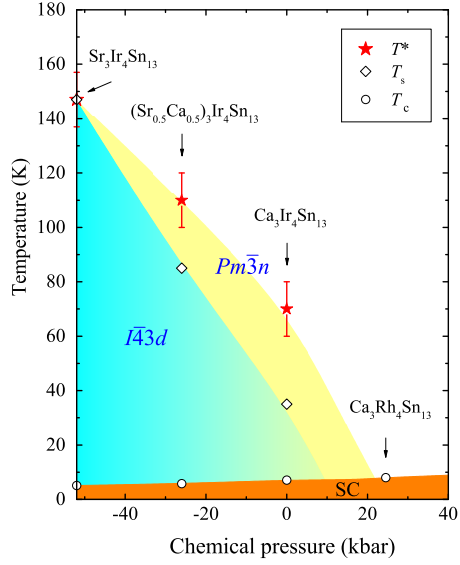}
\centering
\caption{(color online) 
 The phase diagram of (Sr$_{1-x}$Ca$_{x}$)$_{3}$Ir$_{4}$Sn$_{13}$ ($x$ = 0, 0.5, 1) and Ca$_{3}$Rh$_{4}$Sn$_{13}$. $T^*$ and $T_{\rm s}$ are obtained from NMR spectra line width,  $1/T_{1}T$ and the Knight shift. For Sr$_{3}$Ir$_{4}$Sn$_{13}$,  the anomaly above $T_{\rm s}$ can be identified from $1/T_{1}T$ but less clear in FWHM and $K$, thus we plot $T^*$ = $T_{\rm s}$ $\pm$ 10 K. The yellow region is a crossover rather than a new phase. The correspondence between pressure and the composition is inferred from Ref. \cite{Grosche1,Grosche2}.
\label{phasediagram13}}
\end{figure}
In Fig.~\ref{phasediagram13}, we display the phase diagram of (Sr$_{1-x}$Ca$_{x}$)$_{3}$Ir$_{4}$Sn$_{13}$ ($x$ = 0, 0.5, 1) and Ca$_{3}$Rh$_{4}$Sn$_{13}$. The lattice constant of (Sr$_{1-x}$Ca$_{x}$)$_{3}$Ir$_{4}$Sn$_{13}$ shrinks linearly upon substituting Sr with Ca. Therefore, increasing calcium content $x$ is equivalent to applying a hydrostatic pressure ($P$).\cite{Grosche1} We note that Hu $et$ $al$ have also scaled (Sr$_{1-x}$Ca$_{x}$)$_{3}$Ir$_{4}$Sn$_{13}$ and (Sr$_{1-x}$Ca$_{x}$)$_{3}$Rh$_{4}$Sn$_{13}$ in one phase diagram.\cite{Hu} When a pressure of 25.7 kbar is applied to Sr$_{3}$Ir$_{4}$Sn$_{13}$, $T_{\rm s}$ and $T_{\rm c}$ become almost the same with (Sr$_{0.5}$Ca$_{0.5}$)$_{3}$Ir$_{4}$Sn$_{13}$. Such phenomenon also occurs between (Sr$_{0.5}$Ca$_{0.5}$)$_{3}$Ir$_{4}$Sn$_{13}$ and Ca$_{3}$Ir$_{4}$Sn$_{13}$ .\cite{Grosche1} These results suggest that the change in $x$ ($\Delta x$ = 1) corresponds to a change in pressure of $\Delta P$ = 52 kbar. The $T_{\rm s}$ extrapolates to zero temperature at $P$ = 18 kbar in Ca$_{3}$Ir$_{4}$Sn$_{13}$.\cite{Grosche1} In the (Sr$_{1-x}$Ca$_{x}$)$_{3}$Rh$_{4}$Sn$_{13}$ series, $T_{\rm s}$ extrapolates to zero temperature at $x$ = 0.9, or equivalently, at $P$ = -6.8 kbar  relative to Ca$_{3}$Rh$_{4}$Sn$_{13}$.\cite{Grosche2}
Based on these results, 
we obtain the relative pressure for our samples with  respect to Ca$_3$Ir$_4$Sn$_{13}$, and construct a phase
diagram as shown in Fig.~\ref{phasediagram13}.

As the pressure increases, $T_{\rm s}$ and $T^*$ are suppressed while 
$T_{\rm c}$ increases slowly, which means that there exists a competition between the structural phase transition and superconductivity. Similar phase diagram has been seen in other systems such as LaPt$_{2-x}$Ge$_{2+x}$,\cite{LaPt2Ge2} where $T_{\rm c}$ increases from 0.41 K to 1.95 K and $T_{\rm s}$ decreases from 394 K to 50 K. Note that the Knight shift and thus the electronic DOS  remains $T$-independent at low temperatures, while its absolute value increases from Sr$_{3}$Ir$_{4}$Sn$_{13}$ to Ca$_{3}$Rh$_{4}$Sn$_{13}$. Therefore, the increase of DOS maybe partly responsible for the increase of $T_{\rm c}$.

Ca$_{3}$Rh$_{4}$Sn$_{13}$ is located near the end point of the $T^*$ curve, and its superconducting transition
temperature $T_{\rm c}$ = 8 K is close to the highest value of this class of materials under chemical or physical pressures.
In cuprates and Fe-based superconductors,  $T_{\rm c}$  has a close connection to magnetic fluctuations or
structural/orbital fluctuations. Although it is not clear at the moment how the antiferromagnetic spin fluctuations found in this work is related to the structural phase transition,
the antiferromagnetic spin fluctuations may also contribute to the  increase of $T_{\rm c}$. In fact, a systematic change of the Korringa ration S is found as the chemical pressure is increased, as seen in  Fig.~\ref{CaRhSnKorbKorringa} (b). This is a direction needs to be explored in the future.
\subsection{Superconducting properties}
\begin{figure}[h]
\includegraphics[width= 9cm]{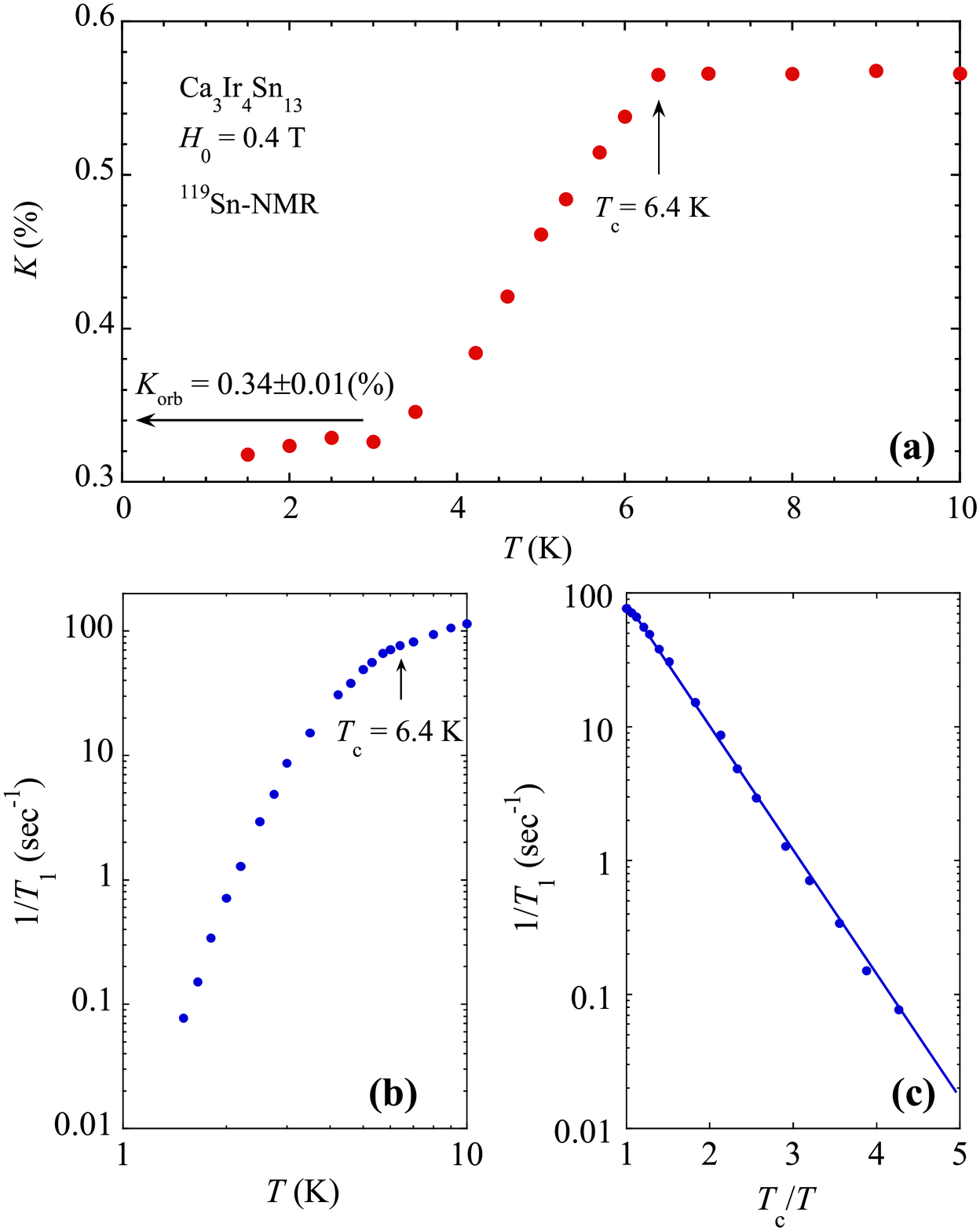}
\centering
\caption{ (color online) 
(a) The temperature dependence of the Knight shift for Ca$_{3}$Ir$_{4}$Sn$_{13}$ at $H_{0}$ = 0.4 T. The arrow indicates $T_{\text{c}}$. (b) The temperature dependence of $1/T_{1}$ below 10 K. (c) The temperature dependence of $1/T_{1}$ below $T_{\rm c}$ in semilogarithmic coordinate. The solid line represents the relation $1/T_{1} \varpropto {\rm exp}(-\Delta/k_{\rm B}T)$
\label{KnightshiftT1T1fit}}
\end{figure}
In this section, we discuss the properties of Ca$_{3}$Ir$_{4}$Sn$_{13}$ in the superconducting state. In order to minimize the effect of the external field,  we have performed NMR measurements for the Sn(2) site under a low field of $H_{0}$ = 0.4 T that is much smaller than $H_{\rm c2} \approx$ 7 T. The $T_{\rm c}$ is   6.4 K at $H_{0}$ = 0.4 T. The $^{119}$Sn Knight shift below $T_{\rm c}(H)$ is shown in Fig.~\ref{KnightshiftT1T1fit} (a). At low temperature of 1.5 K , the Knight shift approaches a value 0.32 $\%$, which is very close to $K_{\rm orb}$ = 0.34 $\pm$ 0.01$\%$ obtained from Fig.~\ref{CaRhSnKorbKorringa} (a) for Ca$_{3}$Rh$_{4}$Sn$_{13}$. A small excess reduction at $T$ = 1.5 K (by about 0.02\%) could be due to the diamagnetism of the vortex lattice.\cite{ZhengPRL} The result indicates that the spin susceptibility vanishes completely at $T$ = 1.5 K, which indicates that the Cooper pairs 
are in the spin-singlet state.

Figure~\ref{KnightshiftT1T1fit} (b) shows the temperature dependence of $1/T_1$. There are two noticeable features. One is that $1/T_1$ decreases exponentially over three decades with decreasing temperature. The other is that $1/T_1$ shows no Hebel-Slichter coherence peak just below $T_{\rm c}$. The lack of a Hebel-Slichter coherence peak was previously reported in filled-\replaced{skutterudite}{Skuterdite} PrOs$_4$Sb$_{12}$.\cite{Kotegawa}
To see the decay more intuitively, we replot $1/T_1$ in a semilogarithmic scale shown in
Fig.~\ref{KnightshiftT1T1fit} (c). As can be seen, a straight line of the relation  $1/T_{1}\propto {\rm
exp}(-\Delta/k_{\rm B}T)$ fitted the data well down to 1.5 K, where $\Delta$ and $k_{\rm B}$ denote the
superconducting energy gap at $T$ = 0 and the Boltzmann constant, respectively. The fitting parameter 2$\Delta$ = 4.42$k_{\rm B}T_{\rm c}$ was obtained, which is larger than the BCS gap size $2\Delta$ = 3.53$k_{\rm B}T_{\rm c}$.
Our result is different from an earlier NMR measurement on Ca$_{3}$Ir$_{4}$Sn$_{13}$ where  a small Hebel-Slichter peak was claimed.\cite{NMR1} The large superconducting energy gap suggests strong-coupling superconductivity \added{which is consistent with the muon spin rotation measurements and specific heat experiments.\cite{musr,musr1,Kenzelmann}} Our result is similar to that of the Chevrel phase superconductor TlMo$_{6}$Se$_{7.5}$,\cite{TlMoSe} in which NMR experiments also revealed a large gap£¬yet with no coherence peak due to the strong phonon damping that suppressed the coherence peak below $T_{\rm c}$.\cite{Allen}
\section{Conclusion}
We have grown single crystals of (Sr$_{1-x}$Ca$_{x}$)$_{3}$Ir$_{4}$Sn$_{13}$ ($x$ = 0, 0.5, 1) and
Ca$_{3}$Rh$_{4}$Sn$_{13}$, and performed electrical resistivity and $^{119}$Sn NMR measurements. In the normal
state, we found an anomaly at  $T^*$ above the structural phase transition temperature  $T_{\rm s}$ in
(Sr$_{1-x}$Ca$_{x}$)$_{3}$Ir$_{4}$Sn$_{13}$ ($x$ = 0, 0.5, 1). The NMR line width increases below $T^*$ and 1/$T_{1}T$ and $K$ \replaced{begin}{begins} to decrease, followed by more distinct changes at $T_{\rm s}$.
None of these anomalies was observed in Ca$_{3}$Rh$_{4}$Sn$_{13}$ that does not undergo a structural phase
transition. Our detailed analysis in (Sr$_{1-x}$Ca$_{x}$)$_{3}$Ir$_{4}$Sn$_{13}$ suggests antiferromagnetic spin fluctuations developing below $T^*$ and becoming more visual below $T\sim$20 K as a possible cause. The increase of $T_{\rm c}$ from Sr$_{3}$Ir$_{4}$Sn$_{13}$ to Ca$_{3}$Rh$_{4}$Sn$_{13}$ can be  partly ascribed to the increase of the electronic DOS, but the antiferromagnetic spin fluctuations may also make a contribution.
Remarkably, the precursory electronic anomaly shares similarity with a  phenomenon under active investigation in the Fe-based high-$T_{\rm c}$ superconductors where a change in the electronic properties expected at $T^*$ occurs already above  $T_{\rm s}$. Therefore, our work sheds lights on other correlated electron systems in a broad context.

In the superconducting state of Ca$_{3}$Ir$_{4}$Sn$_{13}$, the spin susceptibility vanishes at low temperature,
indicating a spin-singlet electron paring.  The spin-lattice relaxation rate decays exponentially with decreasing temperature as ${\rm exp}(-\Delta/k_{\rm B}T)$, which indicates a fully opened energy gap. The large superconducting gap 2$\Delta$ = 4.42$k_{\rm B}T_{\rm c}$ accompanied by a lack of the coherence peak indicates that Ca$_{3}$Ir$_{4}$Sn$_{13}$ is a strong-coupling superconductor.


\end{document}